\documentclass[sn-nature,iicol]{sn-jnl}

\usepackage{graphicx}%
\usepackage{multirow}%
\usepackage{amsmath,amssymb,amsfonts,amsthm,mathrsfs}%
\usepackage[title]{appendix}%
\usepackage{xcolor}%
\usepackage{textcomp}%
\usepackage{subcaption}%
\usepackage{manyfoot}%
\usepackage{booktabs}%
\usepackage{pifont}%
\usepackage{algorithm}%
\usepackage{algorithmicx}%
\usepackage{algpseudocode}%
\usepackage{glossaries}
\usepackage{listings}%

\theoremstyle{thmstyleone}%
\newtheorem{lemma}{Lemma}

\theoremstyle{thmstyletwo}%

\theoremstyle{thmstylethree}%

\makeglossaries
\glsdisablehyper
\newacronym{ANN}{ANN}{Artificial Neural Network}
\newacronym{CRLB}{CRLB}{Cramér–Rao Lower Bound}
\newacronym{MLE}{MLE}{Maximum Likelihood Estimation}
\newacronym{QoE}{QoE}{Quality of Experience}
\newacronym{RAN}{RAN}{Radio Access Network}
\newacronym{RBF}{RBF}{Radial Basis Function}
\newacronym{RSRP}{RSRP}{Reference Signal Recieved Power}
\newacronym{RSRQ}{RSRQ}{Reference Signal Recieved Quality}
\newacronym{UE}{UE}{User Equipment}

\begin{document}

\title[Predicting Drive Test Results in Mobile Networks Using Optimization Techniques]{Predicting Drive Test Results in Mobile Networks Using Optimization Techniques}

\author*[1]{\fnm{MohammadJava} \sur{Taheri}}\email{taheri\_mo96@comp.iust.ac.ir}
\author*[1]{\fnm{Abolfazl} \sur{Diyanat}}\email{adiyanat@iust.ac.ir}
\author[2]{\fnm{MortezaAli} \sur{Ahmadi}}\email{ma.ahmadi@qom.ac.ir}
\author[1]{\fnm{Ali} \sur{Nazari}}\email{nazari\_a17@comp.iust.ac.ir}

\affil*[1]{\orgdiv{Computer Engineering}, \orgname{Iran University of Science and Technology}, \orgaddress{\street{IUST}, \city{Tehran},  \state{Tehran}, \country{Iran}}}

\affil[2]{\orgdiv{Computer Engineering}, \orgname{University of Qom}, \orgaddress{\street{Al-Ghadir}, \city{Qom}, \state{Qom}, \country{Iran}}}

\abstract{%
Mobile network operators constantly optimize their networks to ensure superior service quality and coverage. This optimization is crucial for maintaining an optimal user experience and requires extensive data collection and analysis. One of the primary methods for gathering this data is through drive tests, where technical teams use specialized equipment to collect signal information across various regions. However, drive tests are both costly and time-consuming, and they face challenges such as traffic conditions, environmental factors, and limited access to certain areas. These constraints make it difficult to replicate drive tests under similar conditions. In this study, we propose a method that enables operators to predict received signal strength at specific locations using data from other drive test points. By reducing the need for widespread drive tests, this approach allows operators to save time and resources while still obtaining the necessary data to optimize their networks and mitigate the challenges associated with traditional drive tests.
}

\keywords{Drive Test, Minimal Drive Test, Mobile Networks, Received Signal Strength, Data Prediction, Network Optimization}



\maketitle

\clearpage
\section{Introduction}\label{sec1}
Mobile network operators are constantly working to assess and enhance the performance of their networks. Network optimization involves fine-tuning various parameters to ensure better coverage, increased capacity, and an improved overall user experience. Without continuous optimization, networks can suffer from congestion, weak signal quality, and limited data speeds, all of which negatively impact the user's \gls{QoE}.

From here on, we will refer to network users as \gls{UE}. It is evident that measuring the status of UEs plays a crucial role in the network optimization process. In a mobile network, multiple parameters, such as signal strength and quality, known in 4G networks as \gls{RSRP} and \gls{RSRQ}, can be measured \cite{SAHIN2024104002}. The challenge, however, lies in the fact that many of these parameters cannot easily be accessed by the operator from the network core or \gls{RAN} side. Therefore, it becomes clear that a practical solution is to measure these parameters from the perspective of the end-user, or \gls{UE}.

UE-side data is particularly important as it provides a real-time view of the network's performance from the user's perspective. This data helps identify coverage gaps, detect areas with weak signal quality, and understand user behavior patterns. With this information, operators can make informed decisions about adjusting network configurations, deploying additional infrastructure, or implementing new technologies to enhance network performance. Without accurate user data, optimization efforts would rely solely on theoretical models and assumptions, which may not accurately reflect the real user experience.

\subsection{Motivation}
Mobile network operators use several methods to measure important network parameters, with drive test and minimal drive test being two of the most widely used. Drive test, in simple terms, involves sending teams into the field to gather and analyze network data like signal strength, call quality, and other key metrics. However, even the smallest change in the network setup requires another round of testing, which can be both time-consuming and expensive. To tackle these issues, minimal drive test was introduced. This method uses data sent by UEs, such as in Measurement Reports, to the network, helping to cut down on the need for large-scale field tests. Still, this approach has its own set of challenges, like the fact that it only works when the device is connected and doesn't capture all the needed parameters. Operators must carefully manage these complexities, balancing cost considerations with network optimization and user satisfaction. In short, drive test and minimal drive test are both essential for maintaining high-quality mobile services.

Drive test also faces some real-world challenges. It's costly, requires special equipment, and involves a large team. Plus, there are many areas where conducting a drive test is just not practical. Narrow streets, parks, courtyards, and similar spots are hard to reach or unsuitable for repeated tests. These limitations motivated us to explore how we can predict network performance data in places where running a drive test isn?t feasible. By "predicting data," we mean finding ways to estimate network performance in areas where physically testing isn't possible.

\subsection{Contribution}
\begin{figure}\centering
	\includegraphics[width=\linewidth]{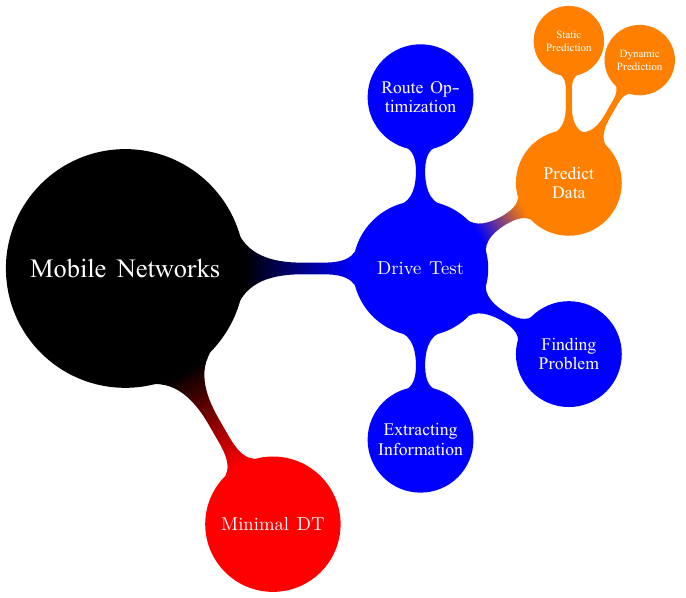}
	\caption{Key Areas in Mobile Network Optimization from a Data Collection Perspective}
	\label{fig:roadMap}
\end{figure}

As shown in \autoref{fig:roadMap}, this paper focuses on drive test in mobile networks. Among the various challenges in this field, our focus is on predicting network performance in areas where data collection is difficult. The aim is to address data gaps in these regions, ensuring they benefit from enhanced network optimization. By predicting data for these locations, we aim to broaden the scope of network optimization and provide a more comprehensive and accurate picture of network performance in all environments. The main innovations presented in this paper are:

\begin{dinglist}{52}
	\item A new approach to estimate path loss parameters for predicting the Reference Signal Received Power (RSRP) in 4G networks, using an optimization-based method with only the location information of the cells.
	\item A novel technique to estimate the standard deviation of noise caused by shadowing in the channel without needing cell location data.
	\item An assessment of our proposed methods using real-world data collected from the Mobile Communication Company of Iran?s network, enabled by a locally developed Android tool called Venus \cite{pesaba2024venus}.
\end{dinglist}

\subsection{Paper Structure}
The rest of the paper is organized as follows: In \S\ref{sec:RelatedWorks}, we review the previous work in this field. \S\ref{sec:ProposedApproach} outlines the system model and assumptions, followed by a detailed explanation of our proposed method. \S\ref{sec:Evaluation} presents an evaluation of the proposed method using real-world data. Finally, in \S\ref{sec:Conclusion}, we provide conclusions and ideas for future research directions.

\section{Related works}\label{sec:RelatedWorks}
In the field of mobile network optimization, there are various methods to enhance network performance and reduce operational costs. As shown in \autoref{fig:roadMap}, network optimization can be achieved through the collection of field data via drive test or through network-side data collection using minimal drive test. When it comes to optimizing drive tests, there are three main approaches: route optimization, network optimization based on data, and data prediction.

The concept of minimal drive test was developed as an alternative to traditional drive test. It leverages data collected by network users during their everyday usage, significantly reducing the need for physical drive tests. By gathering extensive data directly from users, this approach allows for effective network optimization. Studies such as \cite{SKOCAJ2022403,9017353} have explored the use of minimal drive testing to improve network performance and optimize network parameters.

In the realm of drive test optimization, route optimization is a method used before conducting the drive tests. This technique focuses on selecting optimal routes that provide the most valuable data with minimal cost and time. Research papers like \cite{Christofides1986,ARAOZ2009886} offer solutions to the challenges of route planning and optimization.

On the other hand, data-based network optimization involves using data collected from drive tests to improve network configurations. In this approach, once the drive test is completed, the collected data is analyzed, and necessary adjustments are made to enhance network performance. Studies such as \cite{silalahi2021improvement,Peerajing2016Multisector} demonstrate how these changes can lead to improved network service quality.

The primary focus of our research is on data prediction, an approach aimed at estimating network parameters in areas where drive tests have not been conducted. By leveraging existing drive test data, this method predicts parameters such as \gls{RSRP} and \gls{RSRQ}, enabling operators to optimize networks without the need for drive tests in every area. Numerous studies have been conducted in this domain, and we will review some of the most significant ones.

One notable study in this field is presented in \cite{Ojo2020Radial}, which uses drive test data from six base transceiver stations to improve path loss prediction in 4G networks using machine learning. This study illustrates that, unlike traditional models that are rigid and inflexible, machine learning models, such as \glspl{RBF}, offer better accuracy and adaptability, overcoming the limitations of existing models.

In the field of path loss prediction, newer methods have emerged that go beyond the traditional machine learning approaches. \cite{5466252} focus on using \glspl{ANN} to predict path loss for base stations in rural environments. Their findings suggest that a relatively simple ANN model, when trained with drive test data, can outperform traditional models in terms of both prediction accuracy and computational efficiency.

These studies consistently aim to improve the accuracy and efficiency of path loss prediction using drive test data and machine learning. The study in \cite{thrane2018drive} expands on this by utilizing machine learning to predict radio frequency characteristics like RSRP, RSRQ, and signal-to-noise ratio. This research employs a deep neural network trained on data obtained from drive tests, including device locations, base station locations, device-to-station distances, and satellite imagery of the environment, to predict key signal quality metrics for 4G mobile networks. Unlike traditional models that rely heavily on lab-generated data, this approach uses real-world drive test data combined with machine learning to enhance prediction accuracy.

Advanced methods for optimizing drive tests have also been introduced. For instance, \cite{10212575} propose an improved method for predicting \gls{RSRP} using drive test data. This study leverages deep learning techniques and drive test data to significantly increase prediction accuracy. By incorporating supplementary features such as 3D antenna gain and digital elevation models, the research demonstrates superior results compared to traditional methods. The study highlights that combining drive test data with additional environmental information can enhance the accuracy of predictions and the effectiveness of drive test optimization.

Finally, the work presented in \cite{9700950} introduces an intelligent machine learning model for predicting \gls{RSRP} that utilizes drive test data along with advanced machine learning techniques. By employing gradient-boosted trees, this research significantly improves prediction accuracy and robustness across different environments. The paper discusses the challenges associated with feature selection and tuning in these models and offers solutions to enhance model performance. The study shows that by using smarter, hybrid models, more accurate and reliable predictions can be achieved, which are crucial for optimizing mobile networks.

\section{Proposed Approach}\label{sec:ProposedApproach}
We begin by outlining the system model and the assumptions that underpin our approach. Following this, we provide a detailed explanation of the proposed method for predicting drive test data.

\subsection{System Model and Assumptions}
We assume that a drive test has been conducted, but certain locations, due to various constraints or intentionally, have not been covered. The goal of this study is to estimate the network parameters for these uncovered locations, an area referred to as data prediction.

Data prediction can generally be divided into two main categories:
\begin{itemize}
	\item Static Prediction: In this approach, it is assumed that the drive test has been completed, and the aim is to estimate the uncovered locations using various methods.
	\item Dynamic Prediction: This approach involves actively participating in the drive test process in real-time, where the model is continuously updated with new data as it becomes available.
\end{itemize}
To better understand the modeling process and the implementation of the proposed method, the following assumptions have been made:
\begin{enumerate}
	\item The network under study is assumed to be a 4G network. This assumption allows for a focused analysis of the specific challenges and needs of this generation of networks. However, the proposed methods can easily be extended to other network generations.
	\item The focus is on the RSRP parameter. RSRP is a critical indicator of signal quality and plays a key role in evaluating and improving the performance of mobile networks.
	\item It is assumed that the locations and configurations of all base stations in the network are known. This information is essential for creating an accurate and effective model since the base station locations directly impact RSRP prediction results.
	\item The drive test is conducted using a mobile device moving at a speed of less than 40 km/h.
	\item The ellipsoidal coordinate system is used as the reference coordinate system.
\end{enumerate}

\subsection{Proposed Methodology}
\subsubsection{Points Selection}
The first step in the proposed approach is to select suitable points for predicting RSRP values. This process is crucial because the data from these points form the foundation for subsequent calculations and predictions.
\begin{figure}\centering
	\includegraphics[width=\linewidth,page=1]{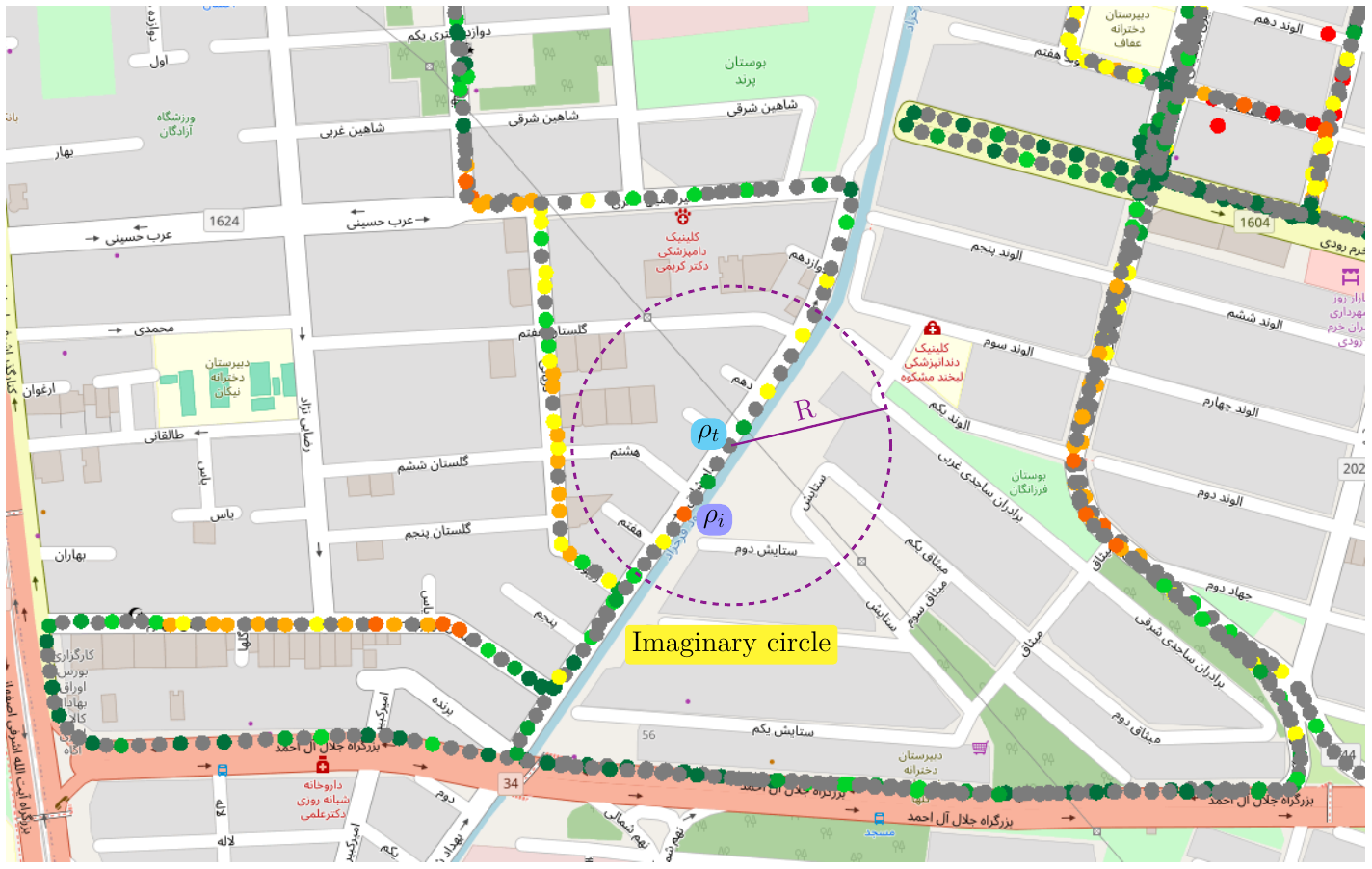}
	\caption{Illustration of a Drive Test: The point $\rho_t$ is plotted on a hypothetical circle with radius $R$.  Points within the circle form the set $\Phi$.}
	\label{fig:choosePoint}
\end{figure}

\autoref{fig:choosePoint} illustrates an example of a drive test. The points shown in gray represent locations where RSRP was not measured, while the colored points indicate locations where RSRP was measured.

Suppose we want to estimate the RSRP value at a point with coordinates
$\rho_t = (x_t,y_t)$
To do this, we draw a circle with radius $R$ around this point. All points within this circle, for which RSRP information is available, are considered in the channel modeling and estimation process. These points inside the circle are denoted by $\Phi$ and are defined as follows:
\begin{equation}
	\Phi = \{\rho_i | d_{it}<R, \text{Has RSRP measurement} \},
\end{equation}
Here, $d_{it}$ represents the distance between $\rho_t$ and $\rho_i$. The radius $R$ of the circle plays a crucial role in prediction accuracy. Increasing the radius allows more points to be included in the calculations, thereby providing the model with more data. However, if the radius is too large, it could reduce prediction accuracy due to the inclusion of points with different environmental and cellular conditions. Conversely, reducing the radius might limit the number of available points, potentially lowering prediction accuracy. However, a smaller radius generally results in more homogeneous conditions among the points, which can increase the accuracy of the results. Therefore, choosing an appropriate radius $R$ is vital for the estimation process.

\subsubsection{Cells Identification and Points Grouping}
Within the set $\phi$, several points have associated RSRP values. These points might be connected to one or more serving cells. The first step here is to identify the serving cells present within this area and then group the points based on the cell to which they are connected.

Consider a point $\rho_i \in \Phi$ that lies within a circle of radius $R$ with a specific RSRP value. Let $C_j$ denote the identifier of the serving cell at that point. We can then divide the points in $\Phi$ into different groups according to their serving cells. Thus, $\Phi_k$ is a subset of $\Phi$ ($\Phi_k\subset\Phi$) where all points are connected to a specific cell:
\begin{equation}
	\Phi_k = \{\rho_i \in \Phi | \rho_i \text{ is connected to } C_j\}
\end{equation}

\begin{figure}\centering
	\includegraphics[width=\linewidth,page=2]{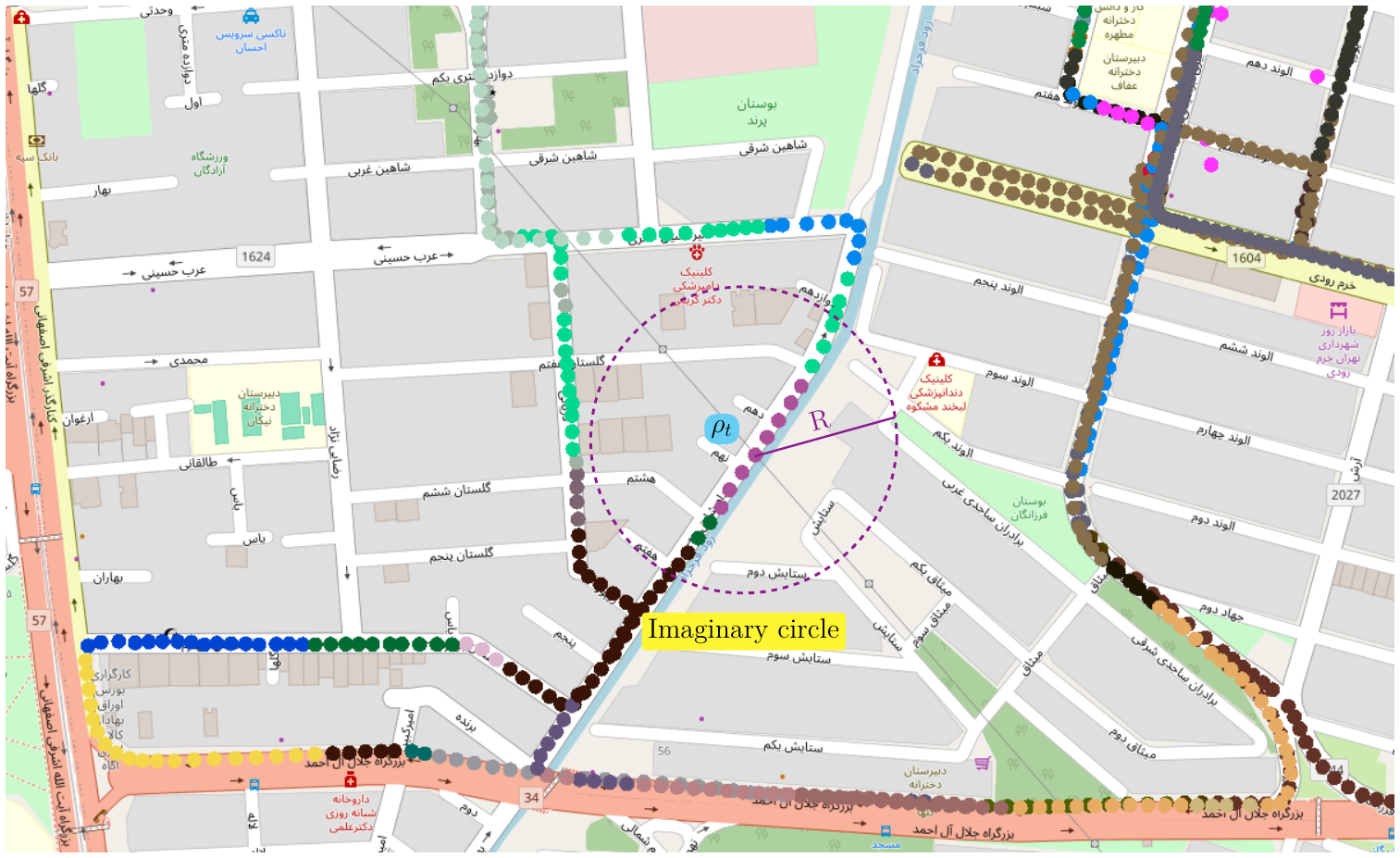}
	\caption{Each color represents a connection to a specific cell. For example, points within the hypothetical circle are connected to four distinct cells.}
	\label{fig:choosePointCell}
\end{figure}
\autoref{fig:choosePointCell} illustrates these concepts. Each color represents a user connected to a different cell. As you can see, in the hypothetical circle centered at point $\rho_t$ the drive test points are connected to four different cells. Consequently, we have four subsets: $\Phi_1$, $\Phi_2$, $\Phi_3$ and $\Phi_4$.

\subsubsection{Path Loss Modeling}
For modeling path loss, we employ the simple yet effective Friis model, widely used in telecommunications due to its minimal parameter requirements and reasonable accuracy. The model is defined as follows:
\begin{equation}
	P_{r} = P_{0} - 10\beta\log_{10}(\frac{d}{d_{0}})\quad \mathrm{[dB]}.
	\label{eq:PiP0beta}
\end{equation}
Here, $P_r$ represents the received power at a distance $d$. The parameter $P_0$ denotes the received power at a reference distance $d_0$. The two key parameters in this model that need estimation are $P_0$ and $\beta$ (the path loss exponent).

First, we identify the points connected to each cell within the area of interest and calculate the distance of these points from their respective transmitters. Various methods for calculating distances in the ellipsoidal coordinate system are discussed in \cite{calculateDistance2015}; we use the following relation in our study:
\begin{align}
		d_{ij} = &R_e \cdot \arccos ( \sin(c \cdot x_i) \sin(c \cdot x_j) 	\label{eq:distancecalcl}\\
& + \cos(c \cdot x_i) \cos(c \cdot x_j) \cos(c \cdot (y_i - y_j)) ),  \nonumber
\end{align}
where $ R_e$ represents the radius of the Earth, which we consider to be $6371000$ meters. Next, using the available data, we aim to estimate the Friis model parameters $P_0$ and $\beta$ by solving an optimization problem. Initially, the measured received power data, along with the locations where the measurements were taken, are organized into a data matrix for each $\Phi_k$, or more precisely, for each individual cell.
\begin{equation}
	\mathrm{Data} = 
	\begin{bmatrix}
		x_{1} & y_{1} & P_{1} \\
		x_{2} & y_{2} & P_{2} \\
		\ldots & \ldots & \ldots \\
		x_{N_k} & y_{N_k} & P_{N_k} \\
	\end{bmatrix}
\end{equation}
where $N_k$ represents the number of measurements for each $\Phi_k$. The parameters $x_i$ and $y_i$ correspond to the latitude and longitude, respectively, and $P_i$ indicates the received power at the 
$i$-th measurement in dBm units.

Our objective is to find the best estimates for the two unknown parameters as described in \eqref{eq:PiP0beta}. If the received power strictly followed the model in \eqref{eq:PiP0beta}, we could estimate the parameters using only two measurements. However, measurements are often not precise and are subject to various sources of error. To model this uncertainty, let's assume that the measurements $P_i$ taken by the user are corrupted by Gaussian noise. This gives us:
\begin{equation}
	P_i =P_0 - \beta \log_{10}d_i +\mathcal{N}(0,\sigma_i), \quad i \in [1,N_k]
	\label{eq:pip0math}
\end{equation}
Here, the noise is assumed to be Gaussian with a mean of zero and standard deviation $\sigma_i$. In other words, we consider only the impact of shadowing noise due to large obstacles obstructing the signal path, ignoring other errors such as user location inaccuracies and thermal noise. Without loss of generality, we also assume that $d_0 = 1$.

Next, we aim to use the \gls{MLE} method to derive an appropriate estimate for the unknown parameters of the problem. It can be shown that in certain cases, MLE provides an optimal solution.
\begin{lemma}
If the noise in equation \eqref{eq:pip0math} follows a Gaussian distribution, then MLE is optimal from the perspective of the \gls{CRLB}, which represents the minimum variance for the parameter we want to estimate.
\end{lemma}
\begin{proof}
Please see \cite[\S\S 11.5.1]{zekavat2019handbook}.

\end{proof}
It is important to note that this rule holds only when the noise in \eqref{eq:pip0math} is Gaussian. In other cases, MLE may not necessarily yield the optimal solution. To estimate the unknown parameters using MLE, we need to construct the likelihood function. If we assume that the measurements are independent of each other, the likelihood function becomes \eqref{eq:fpifrac}, where the parameters in this relation are defined as follows:
\begin{itemize}
	\item $|C_P|$: The covariance matrix of the measurements. It is given by:
	\begin{align}
		|C_P| &= 
		\begin{bmatrix}
			\sigma^2_{1} & 0 & 0 & \ldots & 0 \\
			0    & \sigma^2_{2} & 0 & \ldots & 0 \\
			\vdots & \vdots & \vdots & \ddots & \vdots \\
			0 & 0 & 0 & \ldots & \sigma^2_{N_k} \\
		\end{bmatrix} \nonumber\\
&= \mathrm{diag}(\sigma^2_{1}, \sigma^2_{2}, \ldots, \sigma^2_{N_k})
	\end{align}
	where $\sigma_i^2$ is the variance of the noise for the $i$-th measurement.
	\item The vectors $\overrightarrow{P}$ and $\log_{10}(\overrightarrow{d})$ are defined as:
	\begin{equation}
		\overrightarrow{P} = 
		\begin{bmatrix}
			P_1\\P_2\\\vdots \\P_{N_k}
		\end{bmatrix}\quad \quad 
		\overrightarrow{d} = 
		\begin{bmatrix}
			\log_{10} d_1\\\log_{10} d_2\\\vdots \\\log_{10} d_{N_k}
		\end{bmatrix}
	\end{equation}
	where $d_i$  is the distance between the measurement node and the target node, which can be calculated using \eqref{eq:distancecalcl}. $\overrightarrow{P}$ is essentially the vector of measured powers by the user.
\end{itemize}

By simplifying \eqref{eq:fpifrac} and expanding it from vector notation, we arrive at \eqref{eq:fpifraoverc}. Taking the logarithm of this likelihood function gives \eqref{eq:loglikoloca}:
\begin{figure*}
\begin{equation}
		l(x_t, y_t, P_0, \beta; \overrightarrow{P}) = \frac{1}{\sqrt{(2\pi)^{N_k} |C_P|}}   \
		\times \exp\left( -\frac{1}{2} (\overrightarrow{P} - P_0 + \beta \log_{10} \overrightarrow{d})^T 
		\times |C_P|^{-1} (\overrightarrow{P} - P_0 + \beta \log_{10} \overrightarrow{d}) \right),
	\label{eq:fpifrac}
\end{equation}
\\
\begin{equation}
		l(x_t, y_t, P_0, \beta; P_1, \ldots, P_{N_k}) = \frac{1}{\sqrt{(2\pi)^{N_k} |C_P|}} 
		\times \exp \left( -\frac{1}{2} \sum_{i=1}^{N_k} \left( \frac{(P_i - P_0)}{\sigma_i} \right)^2 
		+  2 \beta \log_{10} (d_i) \right)
	\label{eq:fpifraoverc}
\end{equation}
\\
\begin{equation}
		\log(l(x_t, y_t, P_0, \beta; P_1, \ldots, P_{N_k})) = 
		\log\left(\frac{1}{\sqrt{(2\pi)^{N_k} |C_P|}}\right) 
		\quad - \frac{1}{2} \sum_{i=1}^{N_k} \frac{(P_i - P_0 + \beta \log_{10} d_i)^2}{\sigma_i^2}
	\label{eq:loglikoloca}
\end{equation}
\\
\begin{equation}
		\max \{\log(l(x_t, y_t, P_0, \beta; P_1, \ldots, P_{N_k}))\} \Longrightarrow \quad \\
		\mathrm{min}_{P_0, \beta} \sum_{i=1}^{N_k} \frac{(P_i - P_0 + \beta \log_{10} d_i)^2}{\sigma_i^2}
	\label{eq:minp0}
\end{equation}
\end{figure*}
To apply MLE, we maximize this likelihood function. Since the constant term $\log(\frac{1}{\sqrt{(2\pi)^{N_k} |C_P|}})$ does not affect the maximization process, we can ignore it. Moreover, maximizing a function is equivalent to minimizing the negative of that function. Therefore, we have \eqref{eq:minp0}.
This optimization is feasible when $\sigma_i^2$ is known. In cases where it is challenging to determine $\sigma_i^2$, we can use the Mean Squared Error (MSE) method instead:
\begin{equation}
	\mathrm{min}_{P_0, \beta} \sum_{i=1}^{N_k}(P_i-P_0+\beta \log_{10}d_i)^2
	\label{eq:minpmse}
\end{equation}
In \eqref{eq:minpmse} it is assumed that the shadowing noise is uniform across all measurements belonging to $\Phi_k$. Setting bounds on the optimization variables $P_0$ and $\beta$ can enhance the convergence speed and accuracy of both \eqref{eq:minp0} and \eqref{eq:minpmse}. The bounds are defined as:
\begin{equation*}
	P_{0L} \leq P_0 \leq P_{0H}, \quad\quad \beta_{L} \leq \beta \leq \beta_{H}.
\end{equation*}
These bounds can be expressed using a linear constraint as follows:
\begin{equation}
	\begin{bmatrix}
		+1 & 0 & 0 & 0 \\
		-1 & 0 & 0 & 0 \\
		0 & +1 & 0 & 0 \\
		0 & -1 & 0 & 0 
	\end{bmatrix}\times\begin{bmatrix}
		P_0 \\ \beta \\
	\end{bmatrix} \leq 
	\begin{bmatrix}
		P_{0H} \\ -P_{0L} \\ \beta_H \\ -\beta_L \\
	\end{bmatrix}\quad \Longrightarrow\quad Ax\leq b
	\label{eq:Axbmat}
\end{equation}

Thus, applying the MLE approach to our localization problem results in the following optimization problem:
\begin{subequations}
	\begin{align}
		\mathrm{min}_{x}\quad \quad & \sum_{i=1}^{N}\frac{(P_i-P_0+\beta \log_{10}d_i)^2}{\sigma_i^2} \\
		\mathrm{s.t.}\quad \quad& Ax\leq b
	\end{align}
	\label{eq:MLeoptporblem}
\end{subequations}
Alternatively, in the MSE case, the problem is formulated as:
\begin{subequations}
	\begin{align}
		\mathrm{min}_{x}\quad \quad & \sum_{i=1}^{N}(P_i-P_0+\beta \log_{10}d_i)^2 \\
		\mathrm{s.t.}\quad \quad& Ax\leq b
	\end{align}
	\label{eq:mseoptporblem}
\end{subequations}
where $x=[P_0, \beta]^T$ is the optimization variable vector, and $Ax\leq b$ is an inequality constraint defined as:
\begin{equation*}
	A = \begin{bmatrix}
		+1 & 0 & 0 & 0 \\
		-1 & 0 & 0 & 0 \\
		0 & +1 & 0 & 0 \\
		0 & -1 & 0 & 0 
	\end{bmatrix}
	\quad\quad\quad 
	b = \begin{bmatrix}
		P_{0L} \\ -P_{0H} \\ \beta_L \\ -\beta_H \\
	\end{bmatrix}
\end{equation*}
The solution to the optimization problem \eqref{eq:mseoptporblem} provides the estimated parameters $P_0$ and $\beta$ for each cell, allowing us to use the Friis model to predict signal strength at various points. This model is particularly valuable in real-world applications, especially in complex and large environments, due to its simplicity, computational efficiency, and minimal parameter requirements, enabling accurate modeling of communication channels.

\subsubsection{Shadowing Noise Estimation}

\begin{figure}
	\centering
	\includegraphics[width=\linewidth]{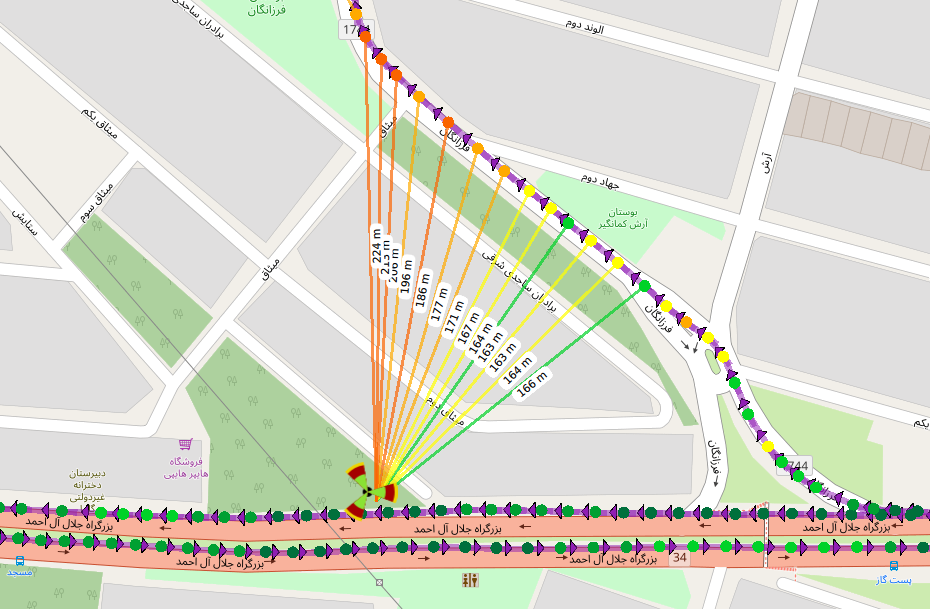}
	\caption{Visualization of Drive Test Points Connecting to Cells}
	\label{fig:dtlocproblem}
\end{figure}

\autoref{fig:dtlocproblem} illustrates how the points recorded in a drive test are connected to their corresponding cell. According to \eqref{eq:pip0math}, the received power is influenced by both path loss and shadowing noise. As discussed earlier, we assume the shadowing effect follows a Gaussian noise model with zero mean and a standard deviation of $\sigma$. In practice, this means that during a drive test, all measurements associated with a particular cell assume that the value of $\sigma$ remains constant across these measurements.

Our first step is to estimate the standard deviation of the shadowing noise, as this parameter plays a critical role in determining the accuracy boundaries of the prediction algorithm. If we know the cell positions, estimating $\sigma$ becomes relatively straightforward: for each hypothetical circle, compute the difference between the measured and estimated RSRP values using parameters derived from \eqref{eq:PiP0beta}, and apply a standard deviation estimator to these differences. However, in the following section, we propose a method to predict $\sigma$ that does not require prior knowledge of the cell locations.

For a set of $N_k$ points belonging to $\Phi_k$, the measured power values can be represented as follows, according to \eqref{eq:pip0math}:
\begin{align}
	P_{1} &= P_0 - 10\beta\log (d_1) + n_1. \nonumber\\
	P_{2} &= P_0 - 10\beta\log (d_2) + n_2. \nonumber\\
	\ldots & \ldots\nonumber\\
	P_{i} &= P_0 - 10\beta\log (d_i) + n_i. \nonumber\\
	\ldots & \ldots\nonumber\\
	P_{N_k} &= P_0 - 10\beta\log (d_{N_k}) + n_{N_k},
	\label{eq:PrM}
\end{align}
where $n_i$ represents samples from a Gaussian distribution with zero mean and standard deviation $\sigma$, denoted as $n_i\sim\mathcal{N}(0,\sigma)$. $P_{i}$ indicates the received power in the 
$i$-th measurement. Several assumptions are considered in these equations:
\begin{itemize}
	\item The parameters $P_0$ and $\beta$ are assumed to be consistent across all measurements for the given cell.
	\item We assume that the values of $P_0$ and $\beta$ are unknown, and due to the lack of information about base station positions, we also cannot compute $d_i$.
\end{itemize}

\begin{figure}\centering
	\includegraphics[width=\linewidth]{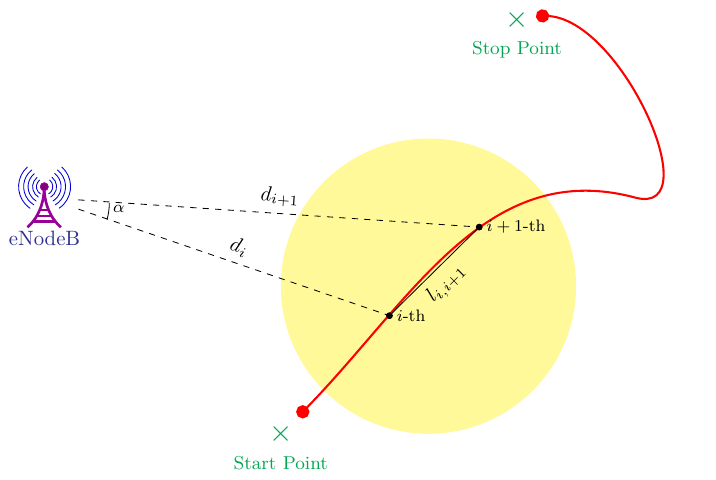}
	\caption{Calculation of Distance Between Consecutive Measurements}
	\label{fig:estimateShadowingNoise}
\end{figure}

Our objective is to estimate the parameter $\sigma$. To achieve this, we draw a circle with radius $R$ around each point in the drive test and subtract \eqref{eq:PrM} for points located within this circle, pairwise. For example, suppose there are three points within the circle; then we have:
\begin{align}
	P_{r1} &= P_0 - 10\beta\log (d_1) + n_1. \nonumber\\
	P_{r2} &= P_0 - 10\beta\log (d_2) + n_2. \nonumber \\
	P_{r3} &= P_0 - 10\beta\log (d_3) + n_3 \nonumber\\
	---&------------\nonumber\\
	P_{r2}-P_{r1} &= 10\beta\log\left(\frac{d_1}{d_2}\right) + n_2-n_1 \nonumber\\
	P_{r3}-P_{r2} &= 10\beta\log\left(\frac{d_2}{d_3}\right) + n_3-n_2 
	\label{eq:pr1diff}
\end{align}
Referring to \autoref{fig:estimateShadowingNoise}, the following theorem can be considered.
Theorem 2: If the displacement between two consecutive measurements is sufficiently small relative to one of them, then $d_{i+1} \approx d_i$.\\
Proof: Consider the triangle depicted in \autoref{fig:estimateShadowingNoise} once more. From trigonometric relationships, we have:
\begin{equation}
	l_{i,i+1}^2 = d_{i+1}^2+d_i^2 - 2d_id_{i+1}\cos\alpha,
	\label{eq:liiid}
\end{equation}
where $l_{i,i+1}$ represents the displacement between two consecutive measurements, and $d_i$ and $d_{i+1}$ are the distances from the target to points $i$ and $i+1$, respectively. If we set $k=\frac{d_i}{d_{i+1}}$, we can simplify \eqref{eq:liiid} to:
\begin{equation}
	\begin{split}
		l_{i,i+1}^2 &= d_{i+1}^2 + d_i^2 - 2d_i d_{i+1} \cos\alpha, \\
		l_{i,i+1}^2 &= d_{i+1}^2 \left( 1 + \left( \frac{d_i}{d_{i+1}} \right)^2 
		- 2\left( \frac{d_i}{d_{i+1}} \right) \cos\alpha \right), \\
		\left( \frac{l_{i,i+1}}{d_{i+1}} \right)^2 &= 1 + k^2 - 2\cos\alpha k.
	\end{split}
\end{equation}

Assuming $l_{i,i+1}$ is much smaller than $d_{i+1}$, the ratio $\frac{l_{i,i+1}}{d_{i+1}}$ approaches zero, leading to:
\begin{equation}
	k^2-(2\cos\alpha) k + 1= 0,
\end{equation}
The solutions to this quadratic equation are:
\begin{equation}
	k = \cos\alpha \pm \sqrt{\cos^2\alpha -1}.
	\label{eq:kcosalp}
\end{equation}

For \eqref{eq:kcosalp} to have a real solution, $\alpha$ must be zero, implying that $k=1$, which means $d_{i+1}=d_i$. In simpler terms, if the displacement between two consecutive measurements is small relative to the distance from the measurement point to the target, then the distances between consecutive measurement points will be approximately equal. This intuitive idea can be easily understood as well.

By leveraging this concept, we can more accurately estimate the shadowing noise standard deviation without needing explicit knowledge of cell locations, enabling more robust and precise modeling in real-world environments.

Given Theorem 2 and \eqref{eq:pr1diff}, we derive the following:
\begin{align}
	P^d_{1}=P_{r2}-P_{r1} &= n_2-n_1 \nonumber\\
	P^d_{2}=P_{r3}-P_{r2} &= n_3-n_2 
	\label{eq:pr1diffconclude}
\end{align}
Since we already know the values of $P_{r1}$ through $P_{rN_k}$, their differences can also be determined. Let's denote this difference by $P^d_{i}$. We can treat the random variable $\mathrm{P}^d$ as the difference between two consecutive received power measurements. This allows us to introduce the following lemma.
\\Lemma 1: The random variable $\mathrm{P}^d$ follows a Gaussian distribution with a mean of zero and a standard deviation of $\sqrt{2}\sigma$.
\\Proof: We know that $n_i\sim\mathcal{N}(0,\sigma)$, and for shadowing noise, the samples are independent over time. Furthermore, if $X$ and $Y$ are independent Gaussian random variables, the difference $X-Y$ is also Gaussian with a mean of zero and a standard deviation of $\sqrt{2}\sigma$. Thus, the random variable $\mathrm{P}^d$ must also follow a Gaussian distribution with these parameters \cite{ross2015first}.

Now that we have a number of samples from the random variable $\mathrm{P}^d$, our goal is to estimate its standard deviation. If successful, we can calculate the shadowing noise standard deviation as:
\begin{equation}
	\sigma = \frac{\sigma_{\mathrm{P}^d}}{\sqrt{2}}.
	\label{eq:sigmafracesi}
\end{equation}

Since we already know the mean ($\mu=0$), the standard deviation $\sigma_{\mathrm{P}^d}$ can be estimated as follows:
\begin{equation}
	\sigma_{\mathrm{P}^d} = \sqrt{\frac{1}{N_k}\sum_{i=1}^{N_k}(P_i^d)^2}
	\label{eq:signamathrm}
\end{equation}
Here, $N_k$ represents the total number of available difference samples, and $P_i^d$ is the difference between the 
$i$-th pair of consecutive measurements. Combining equations \eqref{eq:sigmafracesi} and \eqref{eq:signamathrm}, we obtain:
\begin{equation}
	\sigma = \sqrt{\frac{1}{2N_k}\sum_{i=1}^{N_k}(P_i^d)^2}.
	\label{eq:signamathrdsm}
\end{equation}

Additionally, the $100(1-\alpha)\%$ confidence interval for this estimate is given by:
\begin{equation}
	\begin{split}
		\sqrt{\frac{(N_k-1)\sigma_{\mathrm{P}^d}^2}{2b}} < \sigma < \sqrt{\frac{(N_k-1)\sigma_{\mathrm{P}^d}^2}{2a}},\\ 
		a = \mathcal{X}^2_{(1-\alpha)/2,N_k-1},\quad  b = \mathcal{X}^2_{\alpha/2,N_k-1},
	\end{split}
\end{equation}
where $a = \mathcal{X}^2_{(1-\alpha)/2,N_k-1}$ and $b = \mathcal{X}^2_{\alpha/2,N_k-1}$ represent the values from the Chi-Square distribution with specified degrees of freedom, and $N_k$ is the number of samples.

\section{Evaluation} \label{sec:Evaluation}
In this section, we evaluate the proposed approach using real-world data collected from various drive tests. The experimental setup, data collection process, and the resulting outcomes are analyzed and discussed in detail.

\subsection{Data Collection}
To gather the necessary data, a series of drive tests were conducted across Districts 2 and 4 of Tehran. These tests were carried out using an Android-based application developed by Parto Ertebat Saba \includegraphics[width=5mm]{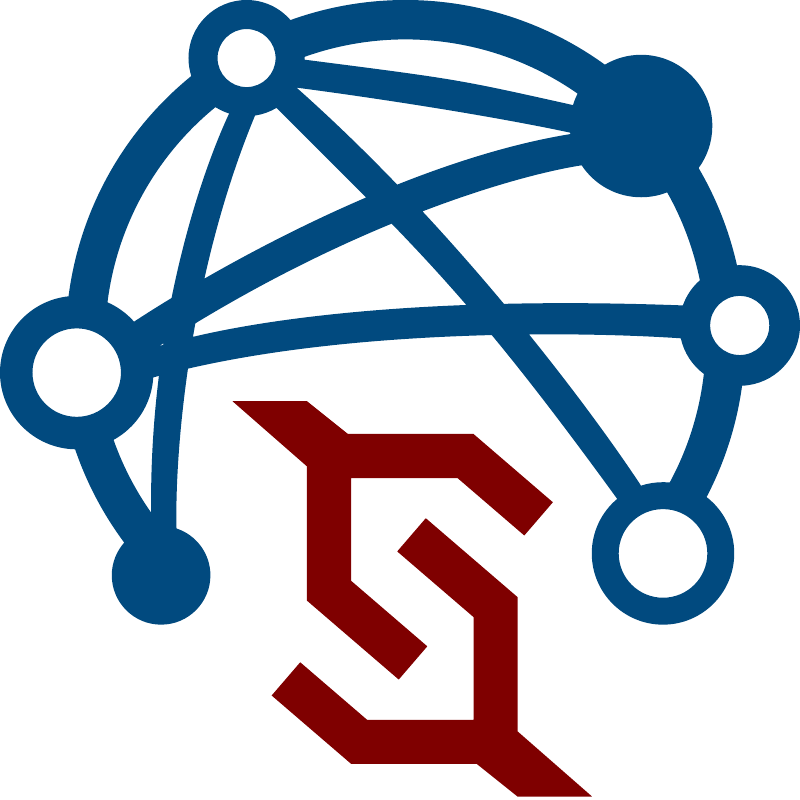}, a knowledge-based company \cite{pesaba2024venus}, with direct involvement from the authors of this paper. The collected data includes signal measurements such as RSRP and RSRQ. In addition to these signal measurements, we also have access to precise location information and identifiers of the serving cells, obtained from available databases. This supplementary information has helped us in grouping the data points and refining the analysis.

The total distance covered during these drive tests exceeded 600 kilometers, with more than 20,000 measurement points accurately recorded. The data collected from these drive tests was used as input for our proposed model, and the evaluation results based on this data are presented in the following sections.

\subsection{Experimental Setup}
\begin{figure}\centering
	\includegraphics[width=\linewidth,page=1]{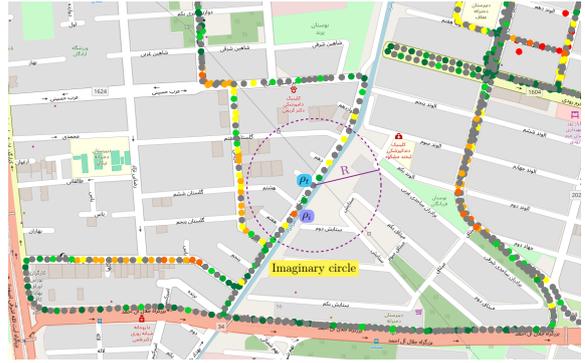}
	\caption{Drive Test Visualization: Gray-colored points indicate where RSRP values were not recorded. These points are excluded from the test dataset.}
	\label{fig:driveTest}
\end{figure}
As shown in \autoref{fig:driveTest}, some points in the collected dataset lack measurements. These points, for which no signal data was recorded, have been excluded from further analysis. The absence of measurements in these areas is due to the fact that the sampling rate was lower than the speed of the moving vehicle. Our main focus is on the points where valid measurements are available to ensure a reliable assessment of the proposed method.

After filtering out the points without measurements, the next step is to estimate the received power at each of the remaining points. Following the proposed method, surrounding points are selected and grouped to be used as inputs for the optimization model, allowing us to accurately estimate the received power.

At this stage, we use the optimization problem described earlier to model the environment?s channel, taking into account shadowing noise. Finally, we estimate the RSRP at each point. This estimation, calculated using the channel model and noise data, determines the received power at each point. To evaluate the proposed method, the estimation error is computed as the difference between the actual RSRP values obtained from the drive test and the estimated values. This error serves as the primary metric for assessing the accuracy of the method.

\subsection{Impact of Parameters on Model Performance}
Several key parameters play a crucial role in modeling the channel and estimating received power. These parameters include:
\begin{itemize}
	\item Radius of the Hypothetical Circle ($R$): This determines which surrounding points are used for estimating the received power at a target location.
	\item Minimum Number of Points Connected to a Cell: This specifies the minimum number of valid measurement points required for a cell to be included in the calculations. If a cell has fewer points, it is excluded from the analysis.
	\item Minimum Distance of Measurement Points from the Serving Cell: This defines the minimum distance a measurement point must be from the antenna to be considered valid for modeling and estimation.
\end{itemize}
These parameters significantly impact the final results, and their effects are analyzed and evaluated in the following sections. We conducted several experiments using the drive test data to assess the influence of these three critical parameters on the performance of our proposed model. For this purpose, we designed two sets of experiments. In the first experiment, we kept the minimum distance of points from the antenna fixed while varying the radius and the minimum number of points connected to a cell. In the second experiment, we fixed the minimum number of points connected to a cell and varied the radius and the minimum distance from the antenna.

To evaluate and compare the model's performance with different parameter settings, we used box plots. These plots allow us to better understand the spread and trends of the measurement errors. Unlike metrics such as the mean, which provide only a general overview of errors, box plots offer a more detailed examination of error distribution and outliers.

\subsubsection{Impact of Number of Points and Circle Radius}
\begin{figure*}
\centering
\begin{subfigure}[t]{0.49\linewidth}\centering
	\includegraphics[width=\linewidth]{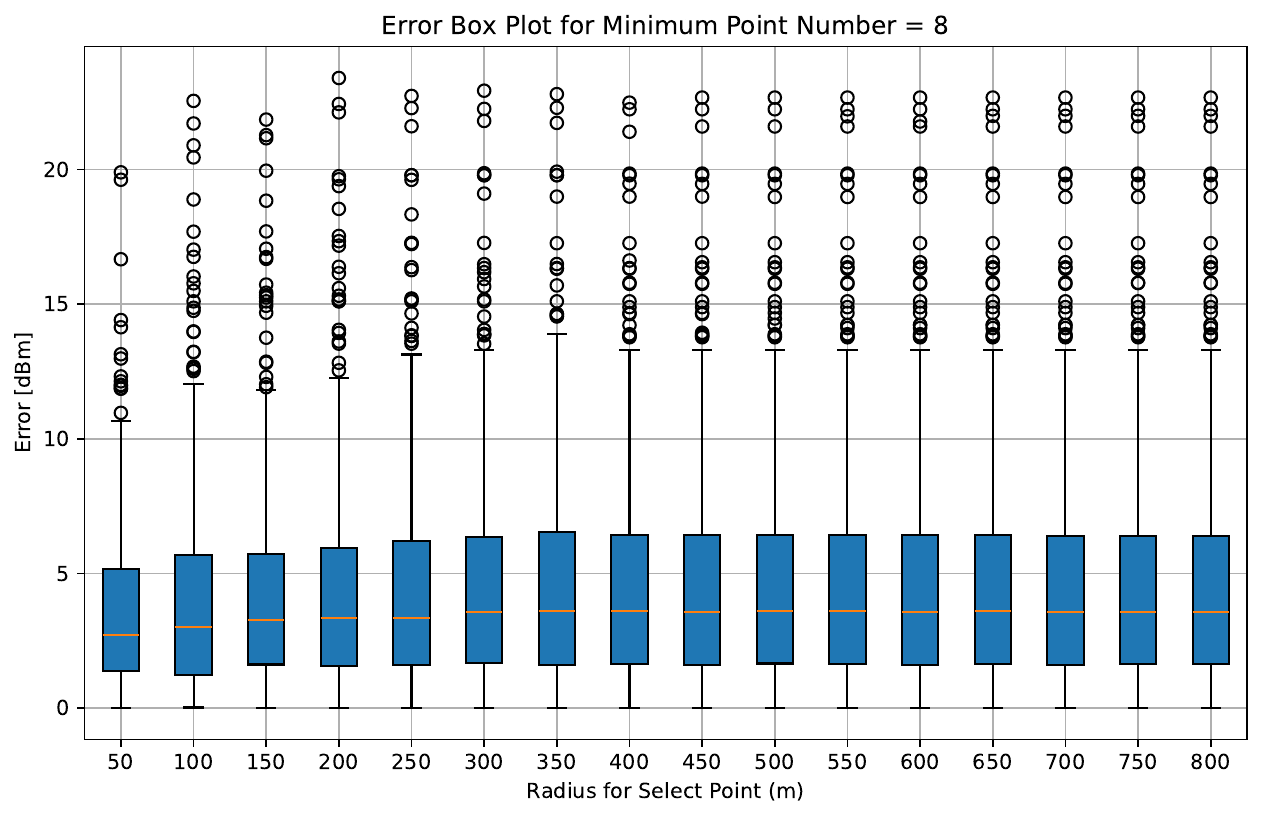}
\caption{}
\end{subfigure}
\begin{subfigure}[t]{0.49\linewidth}\centering
	\includegraphics[width=\linewidth]{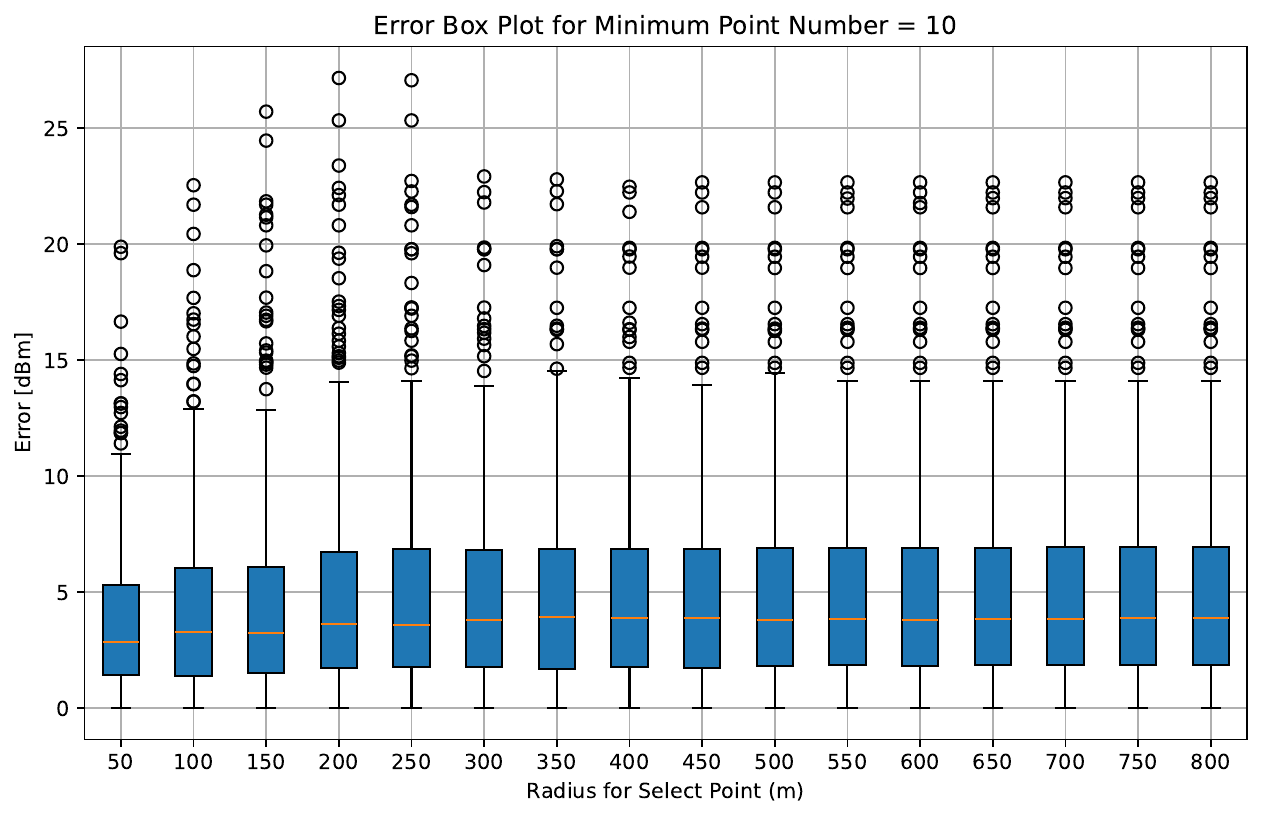}
\caption{}
\end{subfigure}
\\
\begin{subfigure}[t]{0.49\linewidth}\centering
	\includegraphics[width=\linewidth]{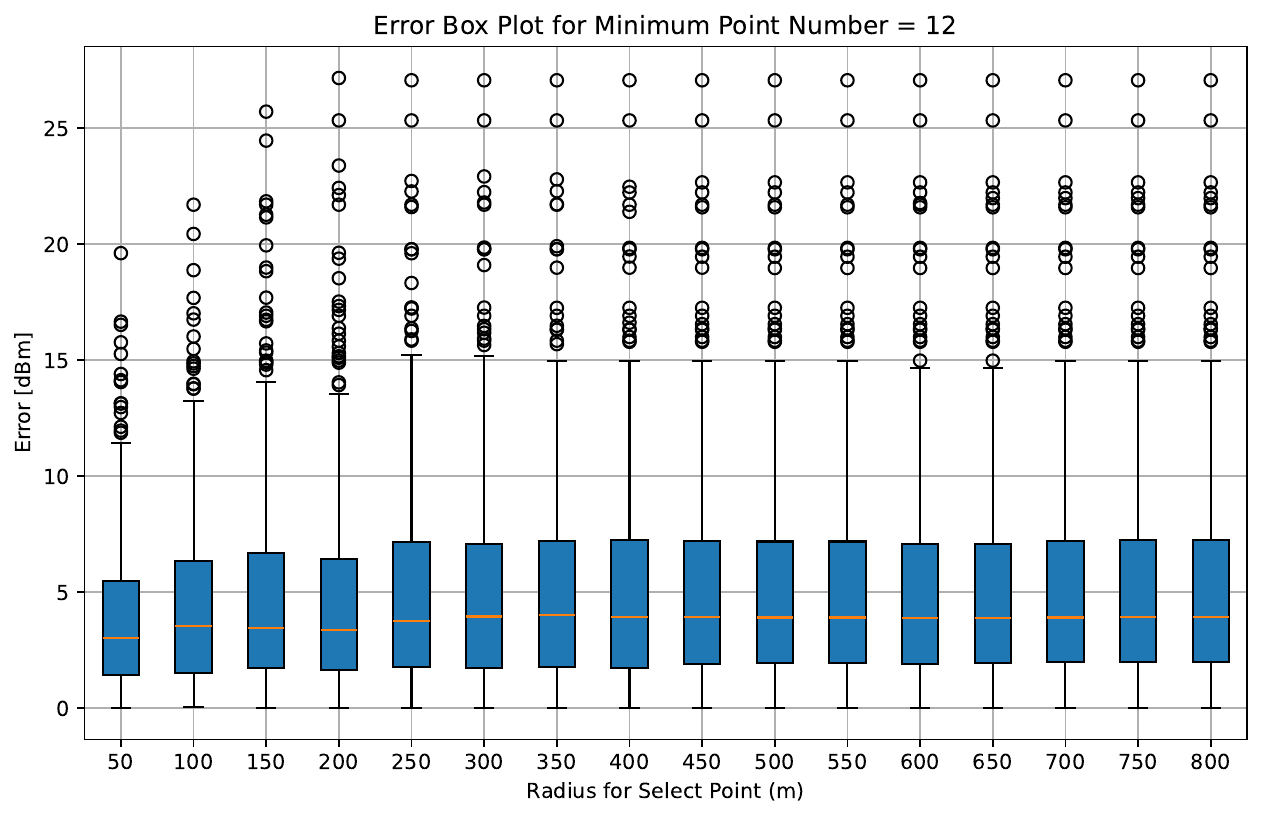}
\caption{}
\end{subfigure}
\begin{subfigure}[t]{0.49\linewidth}\centering
	\includegraphics[width=\linewidth]{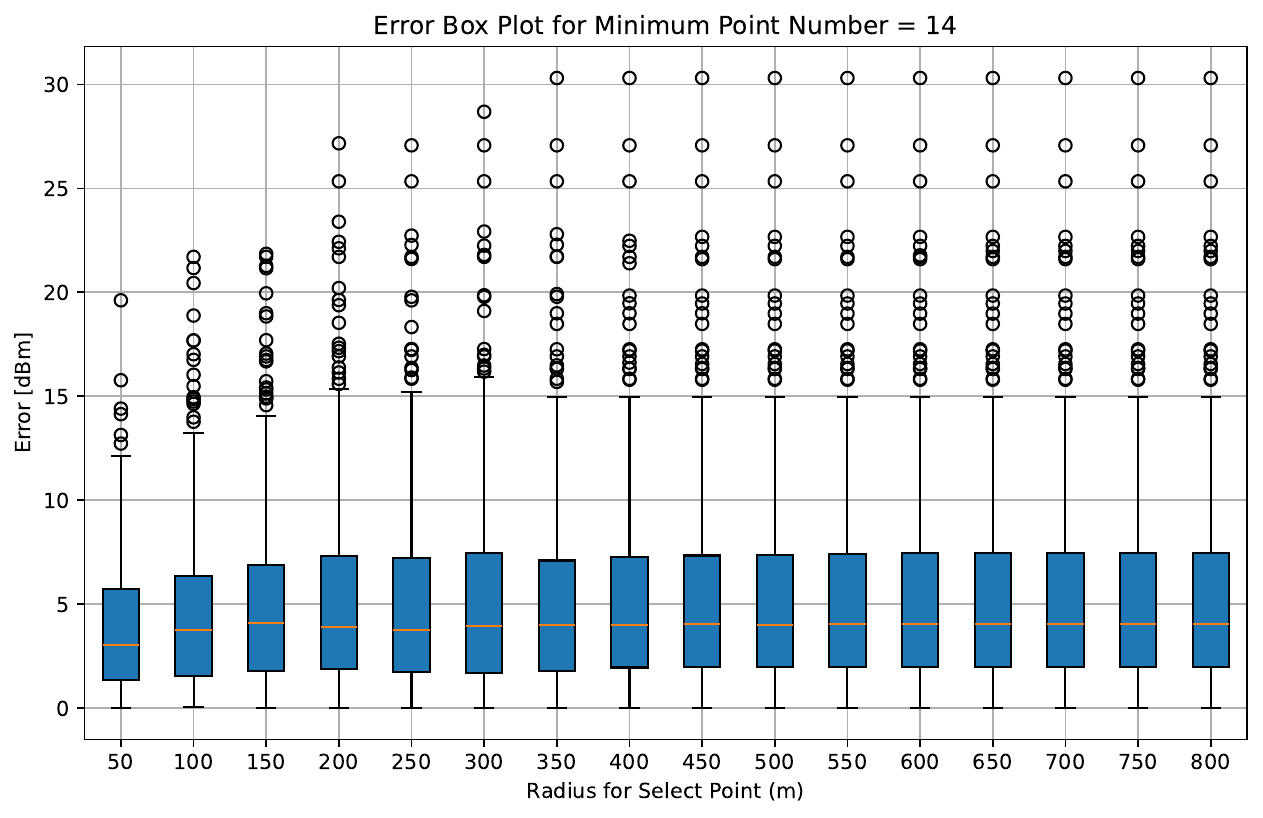}
\caption{}
\end{subfigure}
	\caption{RSRP Parameter Estimation Error for Various Minimum Radius Values and Minimum Number of Points Connected to Cells for Point Counts of 8, 10, 12, and 14}
	\label{fig:minPointNumPlot1}
\end{figure*}

In \autoref{fig:minPointNumPlot1}, each box plot represents a different minimum number of points connected to a cell. Within each plot, separate boxes for various radii display the measured errors. This representation allows us to examine how each model parameter influences the measurement errors and analyze the results.

Based on the experimental results, it is evident that the minimum number of points connected to a cell does not significantly affect the outcomes, as the results remain fairly consistent. Similarly, varying the radius parameter does not result in major differences in the outcomes. However, a radius of 50 meters shows less variation in the RSRP estimation error. This suggests that smaller radii lead to more accurate predictions, likely because points with similar environmental and channel conditions are grouped together. In contrast, as the radius increases, the shadowing noise varies more between measurements, moving the results further from the optimal case.

\subsubsection{Impact of Minimum Distance Parameter}
\begin{figure*}
\centering
\begin{subfigure}[t]{0.49\linewidth}\centering
\includegraphics[width=\linewidth]{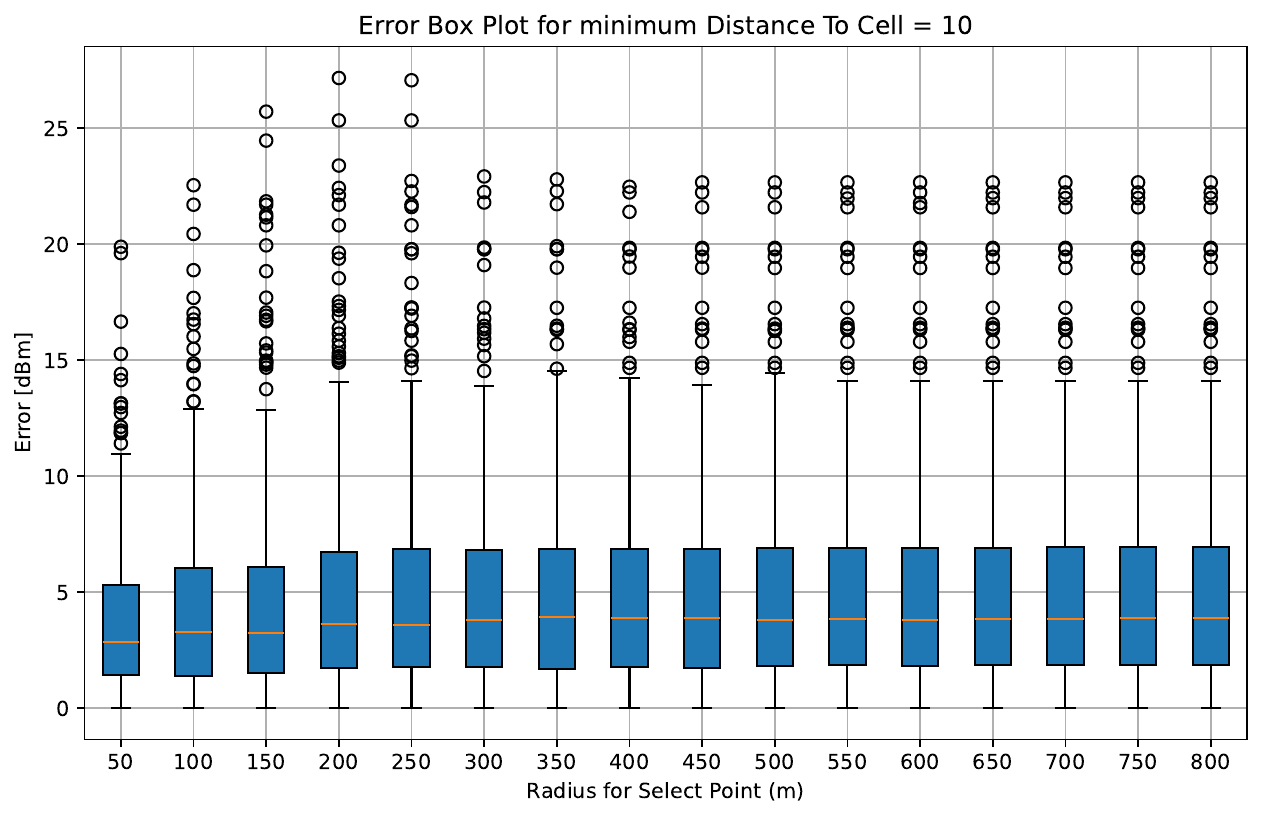}
\caption{}
\end{subfigure}
\begin{subfigure}[t]{0.49\linewidth}\centering
\includegraphics[width=\linewidth]{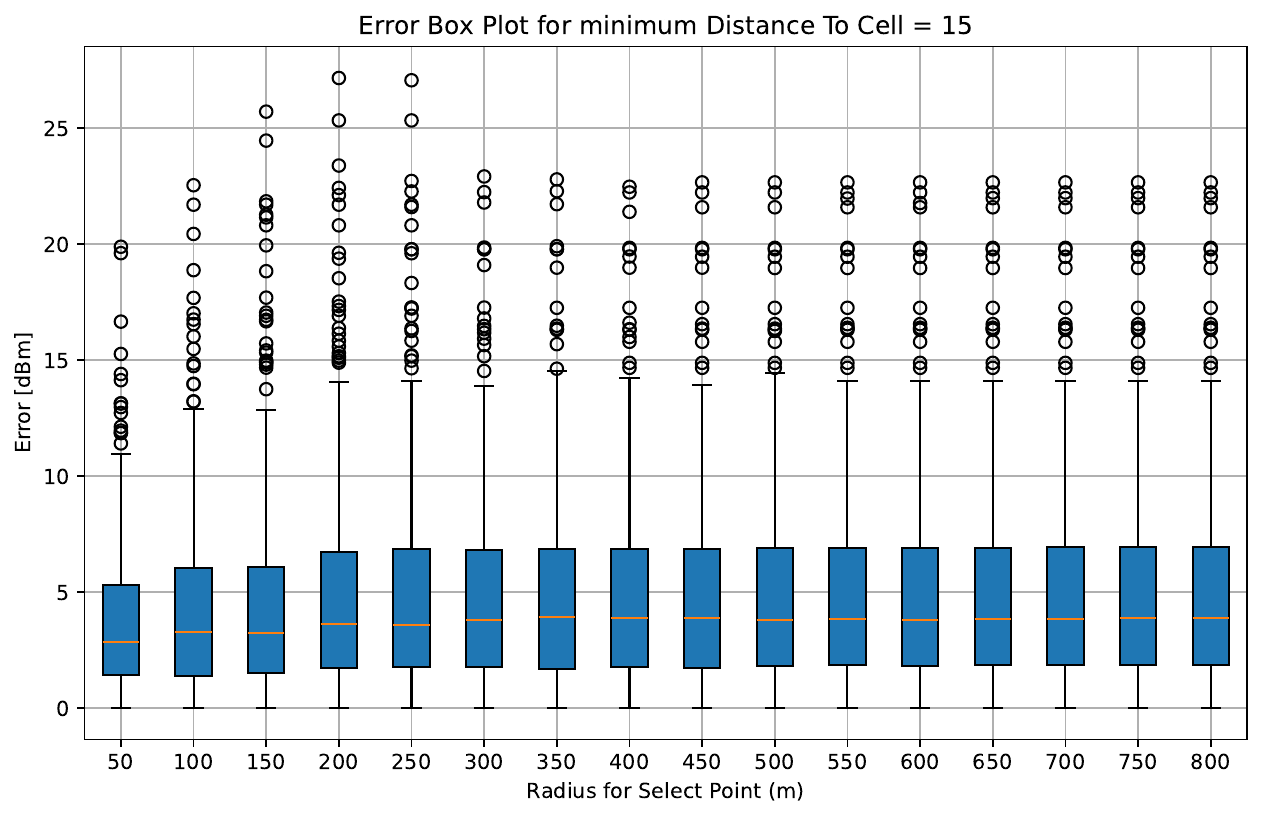}
\caption{}
\end{subfigure}
\\
\begin{subfigure}[t]{0.49\linewidth}\centering
\includegraphics[width=\linewidth]{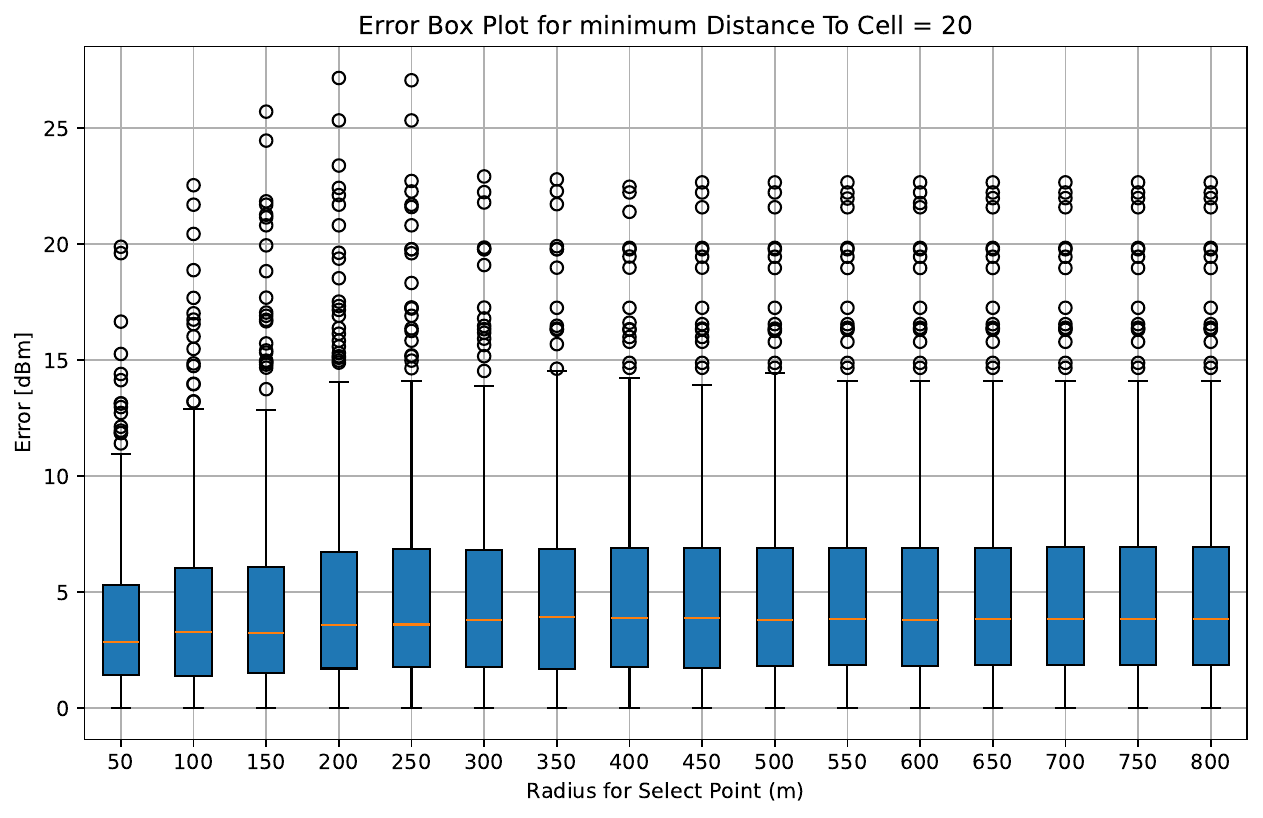}
\caption{}
\end{subfigure}
\begin{subfigure}[t]{0.49\linewidth}\centering
\includegraphics[width=\linewidth]{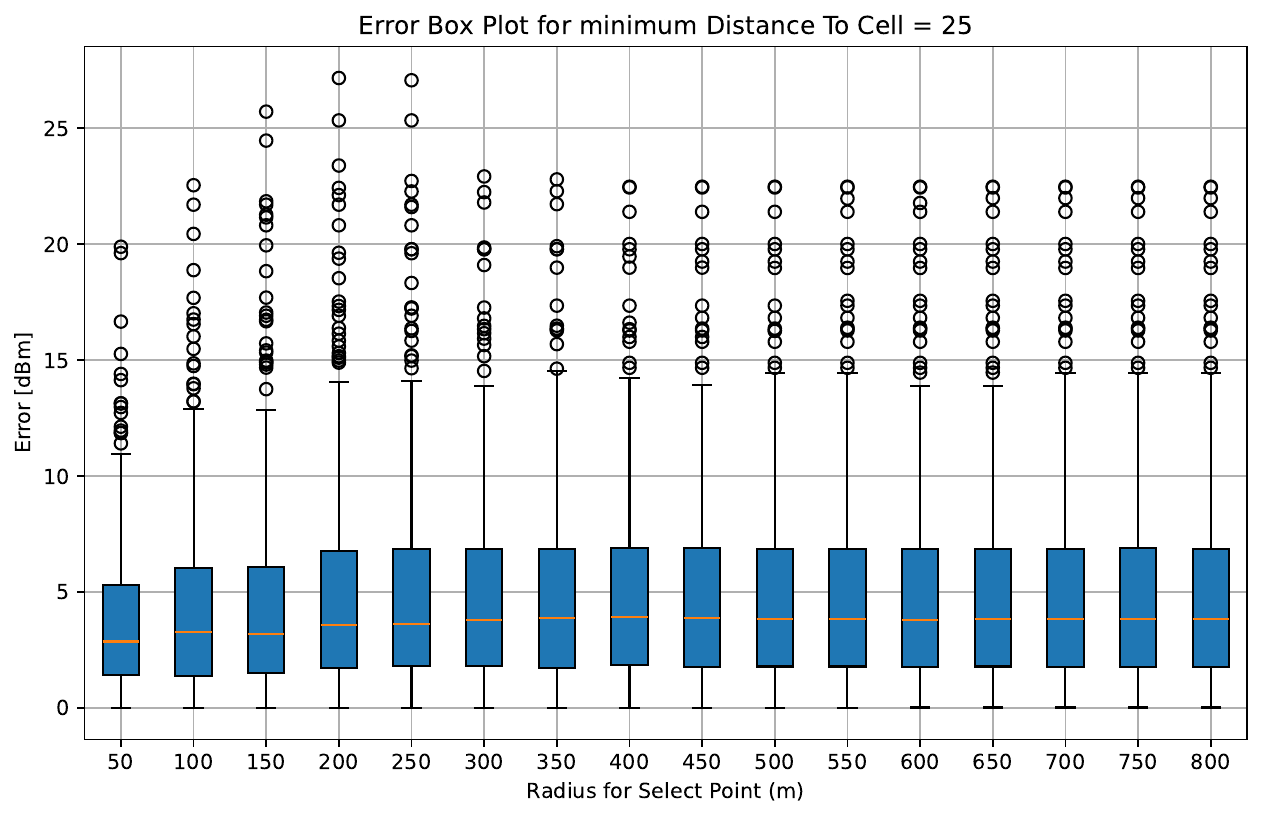}
\caption{}
\end{subfigure}
	\caption{RSRP Parameter Estimation Error for Different Minimum Radius Values and Minimum Distance of Points from the Serving Cell for Minimum Distance of 10m, 15m, 20m, and 25m}
	\label{fig:minDistToCell}
\end{figure*}

As seen in \autoref{fig:minDistToCell}, changing the minimum distance of points from the antenna has a similar lack of impact on model performance as the minimum number of connected points. This is because, when measurement points are located beyond the Fraunhofer distance, \eqref{eq:PiP0beta} remains valid.

\subsubsection{summary}
\begin{figure}
\centering
	\includegraphics[width=\linewidth,page=1]{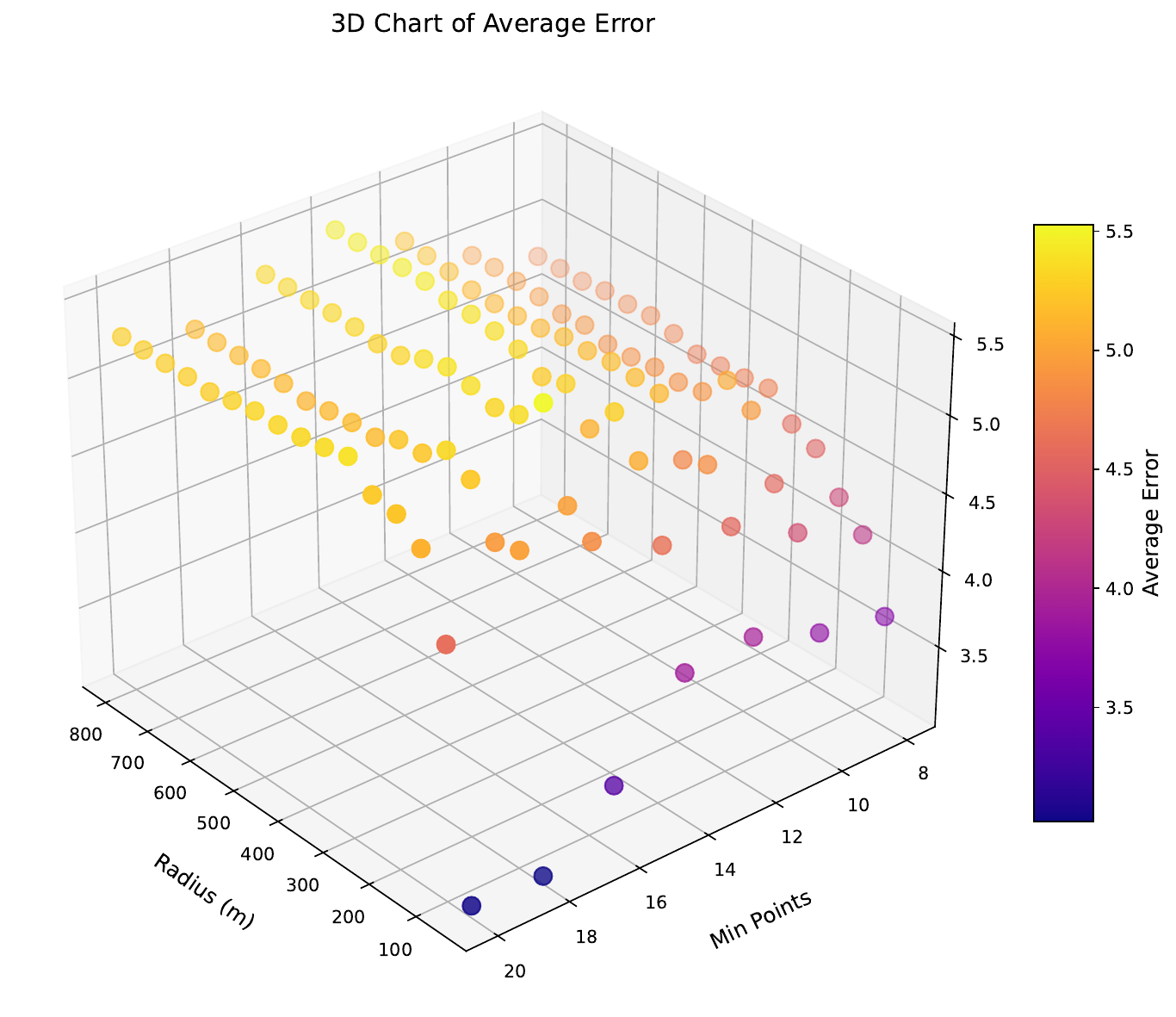}
	\caption{Impact of Minimum Radius and Minimum Connected Points to the Serving Cell on Measurement Error (in dB)}
	\label{fig:3DErrorPlot}
\end{figure}

\autoref{fig:3DErrorPlot} provides a comprehensive overview of how the discussed parameters influence the average error. By analyzing this figure, which illustrates the effects of the radius and minimum number of connected points, we can draw important conclusions regarding the accuracy of the model in predicting RSRP. As shown in the graph, under certain model configurations, the average error is reduced to highly desirable levels. Specifically, with smaller radii and fewer connected points, the model performs better, minimizing prediction error. These results demonstrate the robustness of our model in accurately predicting RSRP. The significant reduction in error highlights the model's effectiveness under optimal conditions.

\subsubsection{Impact of Shadowing Noise on Prediction Error}
\begin{figure*}
\centering
\begin{subfigure}[t]{0.49\linewidth}\centering
	\includegraphics[width=\linewidth]{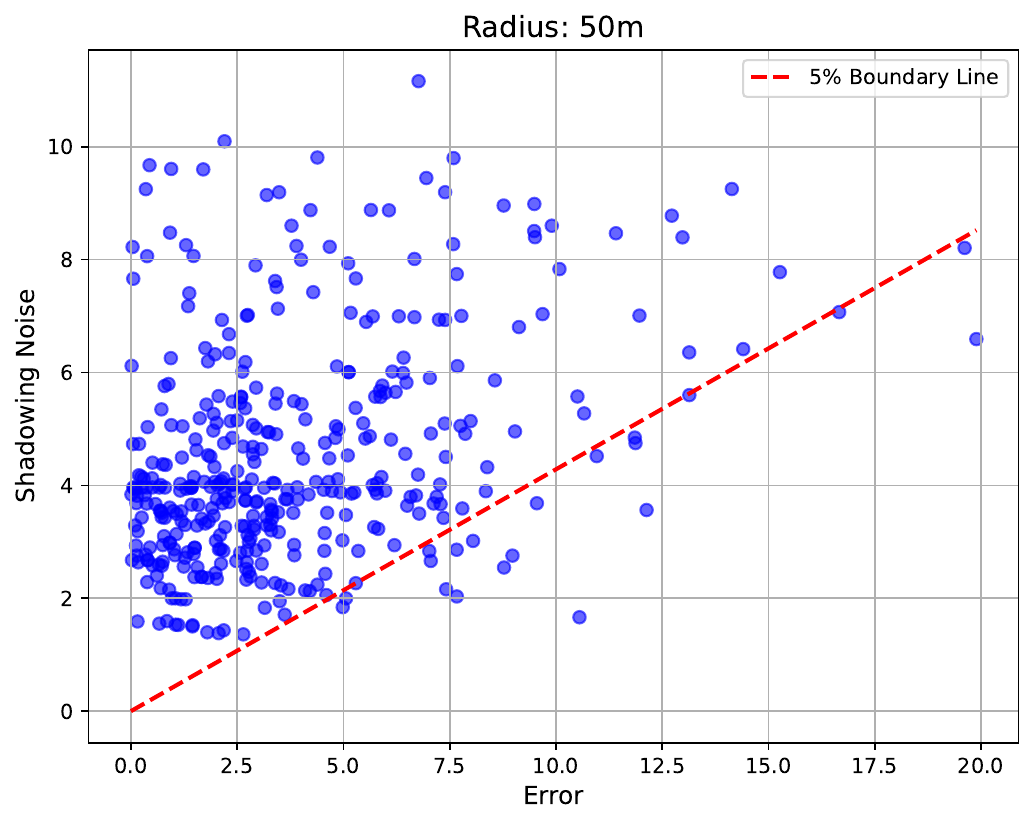}
\caption{}
\end{subfigure}%
~
\begin{subfigure}[t]{0.49\linewidth}\centering
	\includegraphics[width=\linewidth]{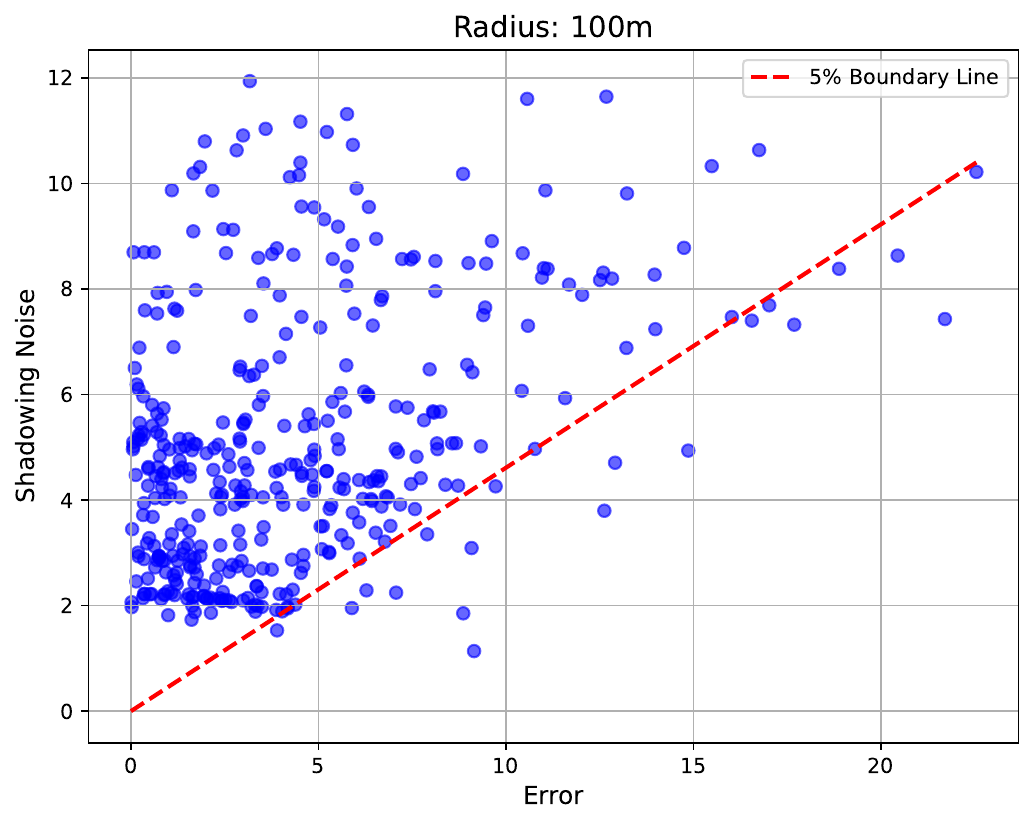}
\caption{}
\end{subfigure}%
\\
\begin{subfigure}[t]{0.49\linewidth}\centering
	\includegraphics[width=\linewidth]{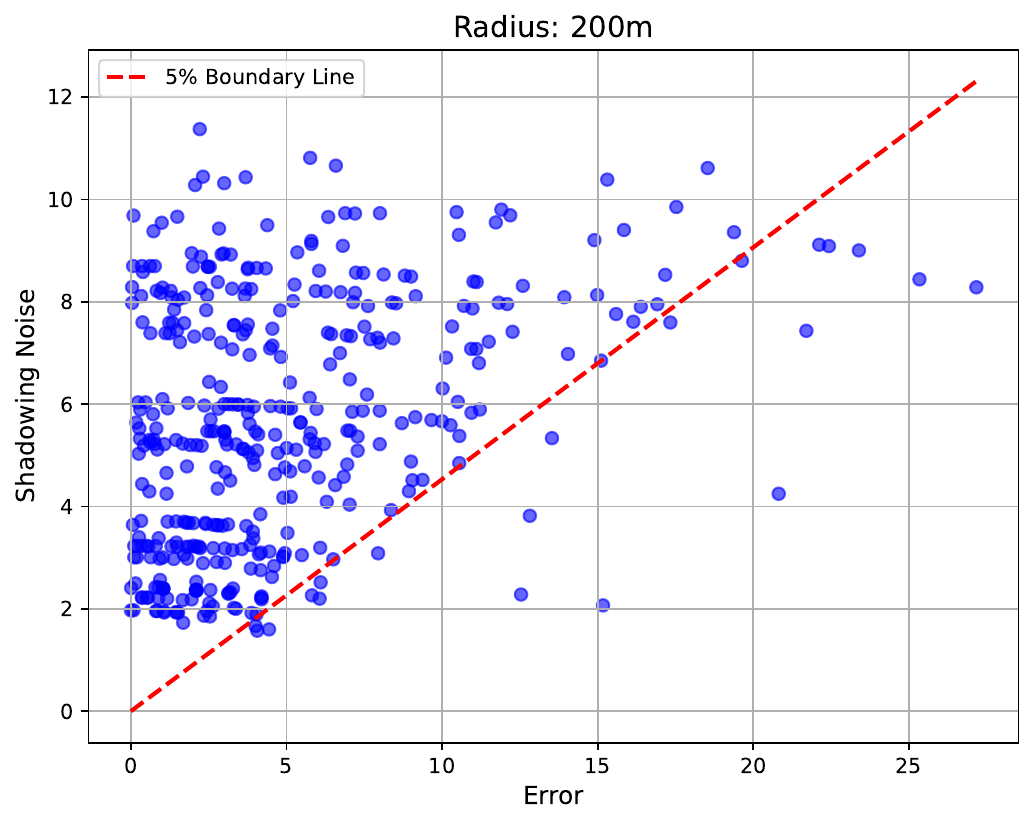}
\caption{}
\end{subfigure}%
~
\begin{subfigure}[t]{0.49\linewidth}\centering
	\includegraphics[width=\linewidth]{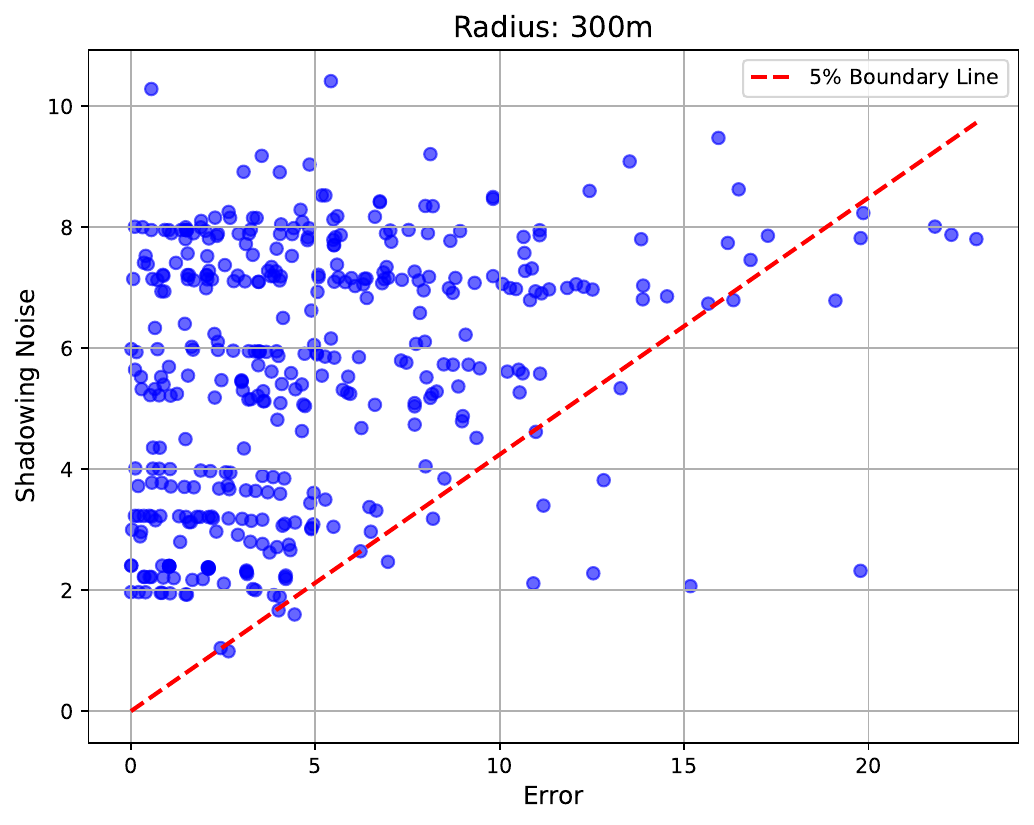}
\caption{}
\end{subfigure}%
	\caption{Scatter plots illustrating the relationship between estimation error and shadowing noise at different radii.}
	\label{fig:Error_Opt_vs_Shadowing_Sigma}
\end{figure*}
In the scatter plots presented in \autoref{fig:Error_Opt_vs_Shadowing_Sigma}, the relationship between prediction error and shadowing noise across different radii is illustrated. A clear pattern emerges from these graphs, indicating that as shadowing noise increases, the prediction error also rises. This trend highlights that shadowing noise plays a crucial role in diminishing the accuracy of the prediction model.

Across all the presented plots, we observe that in regions where the error is low, the shadowing noise is also minimal. In such cases, the model provides highly accurate predictions of the received signal strength, leading to minimal estimation error. However, as the level of shadowing noise increases, the likelihood of larger prediction errors also grows. This direct correlation underscores the significant impact of shadowing noise on the model?s accuracy.

\section{Conclusion}\label{sec:Conclusion}
To optimize their networks, operators rely on accurate data to evaluate various network parameters. One of the primary methods for gathering this data is through drive tests, which collect field data about network performance in different areas. However, conducting drive tests presents several challenges, including high costs, time consumption, and operational difficulties in certain locations. This research aimed to address these challenges by proposing an efficient method for predicting RSRP in areas where drive test data is unavailable.

The proposed approach consists of several key stages, each directly contributing to improved prediction accuracy and reduced reliance on field data collection. First, drive test data critical for prediction was selected based on criteria such as proximity to the prediction point and the minimum distance from the serving cell. Next, we modeled the channel surrounding the target point using mathematical models. At this stage, channel parameters were predicted based on the collected data, allowing us to generate an accurate model of the channel around the prediction point. Following this, the received signal strength at the target point was calculated based on the channel model.

We evaluated the proposed method using real data gathered from multiple drive tests, demonstrating its effectiveness in predicting RSRP while minimizing the need for extensive field data collection.

\bibliography{library.bib}


\end{document}